\documentclass[twocolumn,showpacs,aps,prb]{revtex4}
\usepackage{amssymb}
\usepackage{graphicx}
\usepackage{dcolumn}
\usepackage{bm}
\usepackage{color}

\begin{document}
\preprint{Physical Review B} %

\title{Vortex structures in few-electron quantum dots with spin degree of freedom}

\author{Ning Yang}
\affiliation{%
Department of Physics, and Center for Quantum Information, Tsinghua
University, Beijing 100084, People's Republic of China}
\author{Jia-Lin Zhu}
\email[Electronic address: ]{zjl-dmp@tsinghua.edu.cn}
\affiliation{%
Department of Physics, and Center for Quantum Information, Tsinghua
University, Beijing 100084, People's Republic of China}
\author{Zhensheng Dai}
\affiliation{%
Department of Physics, and Center for Quantum Information, Tsinghua
University, Beijing 100084, People's Republic of China}
\author{Yuquan Wang}
\affiliation{%
Department of Physics, and Center for Quantum Information, Tsinghua
University, Beijing 100084, People's Republic of China}

\date{\today}

\begin{abstract}
The vortex structures and formations of the few-electron states in
quantum dots without the Zeeman splitting are investigated. With
spin degree of freedom, it is noticed that both the choices of probe
electron and the ways to fix the other electrons in conditional
single-particle wave functions affect the display of the vortex
structures and behaviors. Then the vortex transitions in magnetic
fields for the lowest states with different spins are studied. When
the field is not very strong, with the increase of the field, the
vortex number is monotone non-decreasing, and there are absent
states although their angular momenta are in accordance with
transition rules given by the theory of electron molecules.
Different behaviors of the vortices with the change of interaction
range reveal the respective analogies to the vortices of electrons
and quasi-particles in fractional quantum Hall system. The separated
vortices keep apart from the electrons even when the interaction is
screened and such behavior can give an understanding of the absences
of the angular momenta in the transition sequences.
\end{abstract}
\pacs{73.21.La, 71.10.Pm, 73.43.-f}
\maketitle
\section{Introduction}
The formation and distribution of vortices\cite{Toreblad2004} are
important aspects in the studies of many-body systems, such as
superconductors and Bose-Einstein condensates. The vortices can give
the information of the many-body wave functions and the correlations
between particles. The famous Laughlin wave
function\cite{Laughlin1983} whose vortices are concentrated on the
electron is the basis to understand the fractional quantum Hall
effect (FQHE) of two-dimensional electron gas (2DEGs) in strong
magnetic fields. The concentrated vortices keep the electrons far
apart and reduce the short-range interaction most effectively. When
an electron moves around a vortex, the phase of the many-body wave
function will be changed by 2$n\pi$, where $n$ is the order of the
zero. In composite fermion theory\cite{Jain1989} where the FQHE can
be understood in terms of its integer counterpart, the electron
feels the effective magnetic field because the phase change caused
by the vortices partly cancels the Aharonov-Bohm phase caused by the
field.

The quantum dots (QDs) with a few electrons have attracted a lot of
interests in recent years. In magnetic fields, the quantum dots can
be viewed as the precursor of quantum Hall system. So called maximum
density droplet (MDD) which corresponds to the filling factor
$\nu=1$ has been demonstrated by both the experiments and
theoretical studies. Besides these, the electron transport
experiments in magnetic fields have shown various transitions of the
charge distributions.\cite{Oosterkamp1999} This leads to the
increasing interests in investigations of the electronic states and
vortices.\cite{Saarikoski2004, Saarikoski2005} The long-range part
of the interaction is more important in quantum dots. When
$\nu\leq1$, it has been found that the vortices are no longer
concentrated on but bounded around the
electrons.\cite{Tavernier2004, Saarikoski2004} The interaction also
makes the liquid-crystal transition\cite{Yannouleas2002,
Reimann2006, Huang2006} in quantum dots much easier than that in
2DEGs. With the localization of the electrons, the distributions of
vortices become more dispersed to form the vortex clusters.

Due to the Zeeman splitting, the electrons are taken as full
polarized in most theoretical discussions on ground states of QDs in
strong fields. Then the spin degree of freedom can be ignored and is
irrelevant to the studies of vortices. With the improvements of
nanotechnology, it is recently achieved to fabricate the QDs with
negligible Zeeman splitting.\cite{Salis2001, Ellenberger2006} Then
even in the strong magnetic fields, the spin degree of freedom
should be taken into account when the properties of the ground and
low lying states are concerned. The angular momentum transitions of
the ground states and the lowest states with different spins for a
few electrons in quantum dot have been explored by both the theory
of electron molecules\cite{Maksym1995, Maksym2000} and the exact
diagonalization.\cite{Tavernier2003, Tavernier2006} The formation
and redistribution of the vortices with spin degree of freedom are
important aspects of understanding the characters of the electronic
states and the transport measurements in the system without the
Zeeman splitting. In this paper, we study the vortices in
few-electron quantum dots with spin degree of freedom to explore the
transitions of the electronic states in magnetic fields.

\section{Conditional wave function with spin}
The model Hamiltonian of a few-electron quantum dot in the magnetic
field with parabolic confinement and without the Zeeman splitting is
\begin{equation}\label{EQ:Hamil}
H\!\!=\!\!\sum_{i=1}^N {\left [\frac{1}{2m}\left
(\hat{P}_i+e\vec{A}\right )^2\!\!+V(r_i)\right
]}\!\!+\!\!\sum_{i<j}{\frac{e^2}{4\pi\varepsilon|\vec{r}_i-\vec{r}_j|}}.
\end{equation}
where N is the particle number, $\vec{A}$ is the vector potential of
the field, $V(r_i)$ is the confinement of the dot with the strength
equals to $2meV$ in the following discussions, and the last term is
the Coulomb interaction between particles. The effective mass $m$ of
the electron and the static dielectric constant $\varepsilon$ are
respectively $0.067m_e$ and 12.4 for GaAs. The eigenstates $\Psi$ of
the Eq.(\ref{EQ:Hamil}) are obtained by the exact diagonalization.
Without the spin-orbit coupling, such states are the common
eigenstates of the total angular momentum $L$, the total spin $S$
and its z-component $S_z$, so in the following discussions we use
the abbreviation $(L,S,S_z)$ to represent them.

It has been demonstrated that the vortex structures depend on the
range of the interaction between electrons.\cite{Stopa2006} So we
also use the Yukawa screened Coulomb interaction\cite{Ando1982,
Stopa2006}
\begin{equation}
I(r)=\frac{e^2}{4\pi\varepsilon}\frac{\exp(-r/\alpha)}{r}
\end{equation}
instead of the Coulomb interaction to understand the vortex
behaviors in some of the following discussions.

Having got the many-body eigenstates, we employ the conditional
single-particle wave function\cite{Saarikoski2004} with spin degree
of freedom to explicitly show the vortices of the wave functions
\begin{equation}\label{EQ:Cwaf}
\psi_c({\bf r})=\frac{\Psi({\bf r},\sigma^*,{\bf r_2}^*,\sigma_2^*,
\cdots {\bf r}_N^*,\sigma_N^*)}{\Psi({\bf r}^*,\sigma^*,{\bf
r}_2^*,\sigma_2^*, \cdots {\bf r_N^*},\sigma_N^*)}
\end{equation}
where $\sigma_i$ represent the spins of electrons. In the function,
an electron is chosen as the probe electron and other electrons are
pinned at fixed positions. In Eq.(\ref{EQ:Cwaf}), the variables with
asterisk superscript are fixed ones. The phase angle of the
conditional wave function reveals the change of the phase of a
many-body function when an electron moves around another one. Then
the plot of the electron density and the phase of the conditional
wave function gives the picture of vortices in the many-body wave
function.

\begin{figure}[ht]
\includegraphics*[angle=0,width=0.38\textwidth]{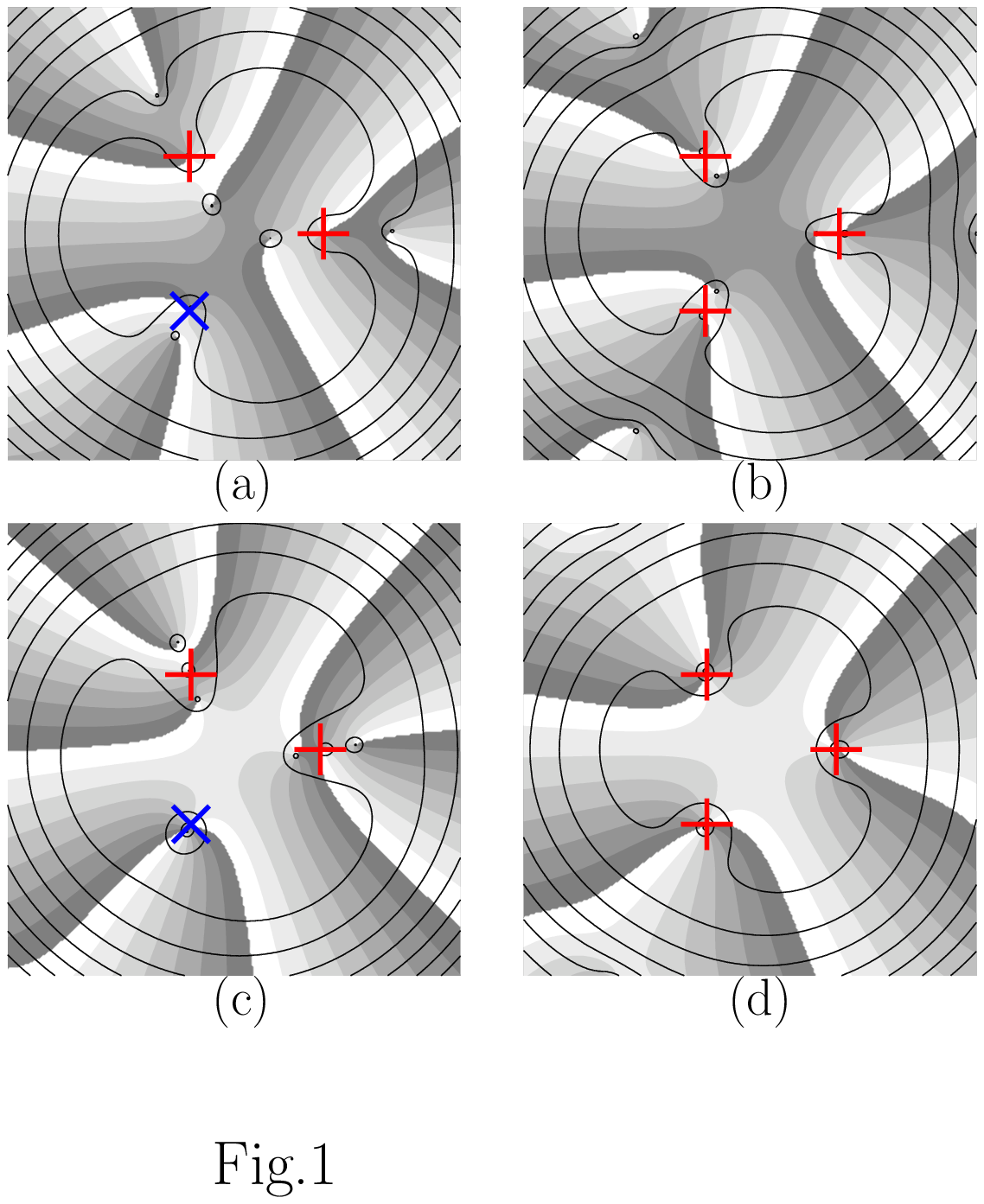}
\caption{\label{FIG:DiffSz} (Color online) The vortices for the
four-electron state (-15,1,1) with a spin-up (left column) and
spin-down (right column) probe electron in conditional
single-particle wave function. The upper and lower rows correspond
to the case where the Coulomb and the screened interactions are
adopted, respectively. The density of the probe electron is plotted
as contours in logarithmic scale. The phase changes from $-\pi$ to
$\pi$ as the shadowing changes from the darkest gray to white. $+$
and $\times$ indicate the positions of pinned spin-up and -down
electrons, respectively.}
\end{figure}

We should present some explanations about the choices of the probe
and pinned electrons. First, when there are asymmetry between the
spin-up and spin-down electrons in some states, the different
choices of the probe electron's spin may result in different
displays of the vortices. In Fig.\ref{FIG:DiffSz}, we present the
vortices of the four-electron state (-15,1,1) as an example. We fix
the pinned electrons at the most probable radius. The state
(-15,1,1) contains three spin-up and only one spin-down electrons.
Then there are two choices of the probe electron and they lead to
different displays of the vortices. If the probe electron is the
spin-up one, it can be seen that there are respectively three and
two vortices around each spin-up and spin-down fixed electron. If
the unique spin-down one is chosen to be the probe electron, only
two vortices around each spin-up electron can be seen. In the case
of the FQHE where the vortices are concentrated on the electrons,
Halperin\cite{Halperin1983} has suggested a set of trial functions
including the spin degree of freedom. The polynomial parts of the
functions have the form
$\prod_{i>j}(z_i-z_j)^{m_+}\prod_{i>j}(\xi_i-\xi_j)^{m_-}\prod_{i,j}(z_i-\xi_j)^n$
where $z$ and $\xi$ are the complex coordinates of the spin-up and
spin-down electrons. It can be found that the orders of the vortices
indeed depend on the spins of the electrons. And from the viewpoint
of the electron with a certain species of spin, only parts of the
vortices are visible. In the case of the quantum dot, although the
vortices are no longer concentrated on the electrons, the number of
the visible vortices can still depend on the spins. Nevertheless,
with appropriate choice of the spin of the probe electron, we can
get useful information about the vortices of a state. In the
following discussions about four-electron states, we mainly focus on
the ones with $S_z=0$ then there are no such difficulty. For
five-electron case with $S_z=0.5$, we chose a spin-up electron as
the probe one.

Another feature should be pointed out here is that there are
vortices in the system which originate from long-range interactions.
And such vortices will move to the infinity if the interactions is
totally screened.\cite{Stopa2006} In Fig.\ref{FIG:DiffSz}b we can
see two such vortices. It can be seen in Fig.\ref{FIG:DiffSz}d that
they will move away when the interaction is screened.

\begin{figure}[ht]
\includegraphics*[angle=0,width=0.38\textwidth]{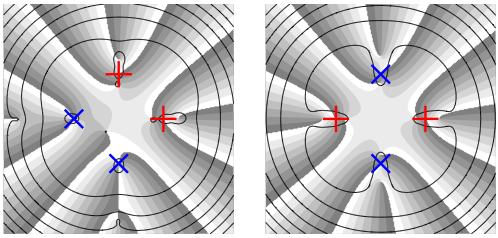}
\caption{\label{FIG:diffpin} The vortex displays of the
five-electron state (-30,0.5,0.5) with the electrons are fixed in
the neighbor (a) and alternative (b) modes, respectively. $+$ and
$\times$ indicate the positions of pinned spin-up and -down
electrons, respectively.}
\end{figure}

Besides the choice of the probe electron, the manner of the pinned
electrons may also affect the vortex displays of some states if
there are more than one way to fix the positions of the electrons
with spins. For instance, if one spin-up electron is selected as the
probe electron for a five-electron state with $S_z=0.5$, and the
remainder electrons are fixed at the vertices of a square, there are
two inequivalent ways to determine the spin of each pinned electron,
i.e. fix the electrons with same spin at neighbor or alternative
positions (we call them neighbor and alternative mode respectively).
Then for the states with some angular momenta, the two ways may give
different vortex numbers. In Fig.\ref{FIG:diffpin} there is an
example of such situation. With two fixation modes, the displayed
vortex numbers of the state (-30,0.5,0.5) are different by one. Such
phenomenon will generally occur when we inspect a higher dimensional
object from a lower dimensional space, i.e. the visibility of an
entity relies on the viewpoint of the projection. It should be
pointed out that such differences only exist in the
non-full-polarized states. For the full-polarized states, the spin
and spatial parts of the wave functions can be separated and the
ways to fix the electrons with different spins do not affect the
display of the vortices. For four electron case, there is only one
way to fix the electrons if they are fixed at the vertices of an
equilateral triangle and of course no such kind of difference exist.
Then in the following counts of the vortices of five-electron states
with $S_z=0.5$, we can simply use appropriate fixation mode which
give larger vortex number for each state. However, when we discuss
the behaviors of the vortices in the subsection III.B, we must
inspect the different displays of the vortices more carefully and we
will recall this topic there.

\section{Disscussion}
\subsection{Energy level structures and spin-dependent vortices}
\begin{figure}[ht]
\includegraphics*[angle=0,width=0.4\textwidth]{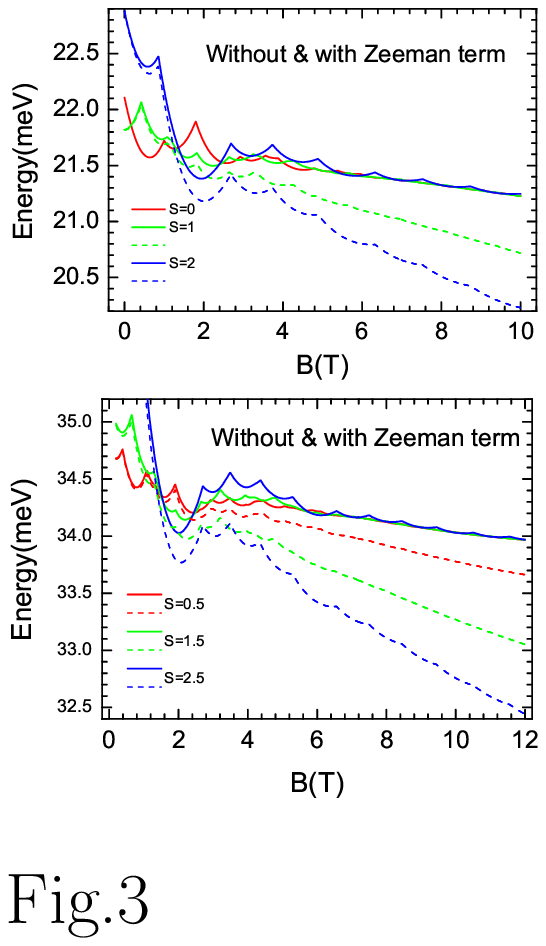}
\caption{\label{FIG:Energy} The energy level of the lowest states
with different spins for four-electron (a) and five-electron (b)
cases as functions of magnetic fields. The solid and dashed lines
correspond to the energies without and with the Zeeman splitting.
The N times of the energy of the first landau level have been
subtracted from the total energy.}
\end{figure}
Before the detailed study of the vortices, we must discuss the
energy level structures of quantum dots briefly. In
Fig.\ref{FIG:Energy} we show typical energy spectra of four and five
electron in magnetic fields obtained from the exact diagonalization.
If the Zeeman splitting is concerned, the ground states of the
few-electron quantum dot are full-polarized with maximum $S_z$ in
strong magnetic fields. With the increase of the field, for the
four- and five-electron case there are angular momentum transitions
of the ground states whose increment between neighbor states equals
to the particle numbers. The allowable angular momenta are so called
``magic number". If the Zeeman splitting can be ignored, the states
with same $L$ and $S$ but different $S_z$ are degenerate. The ground
states even in strong fields no longer need to be full-polarized and
the lowest states with different spins gradually form a narrow band.
In the narrow band, the energies of different spin states are nearly
degenerate. The ground states in the field corresponding to the
fractional filling factor $\nu=1/(2p+1)$ are still full-polarized
although they are multiple degenerate due to different $S_z$. Along
with the gradual formation of the narrow band, there are transition
from the liquid to crystal states.

\begin{figure}[ht]
\includegraphics*[angle=0,width=0.4\textwidth]{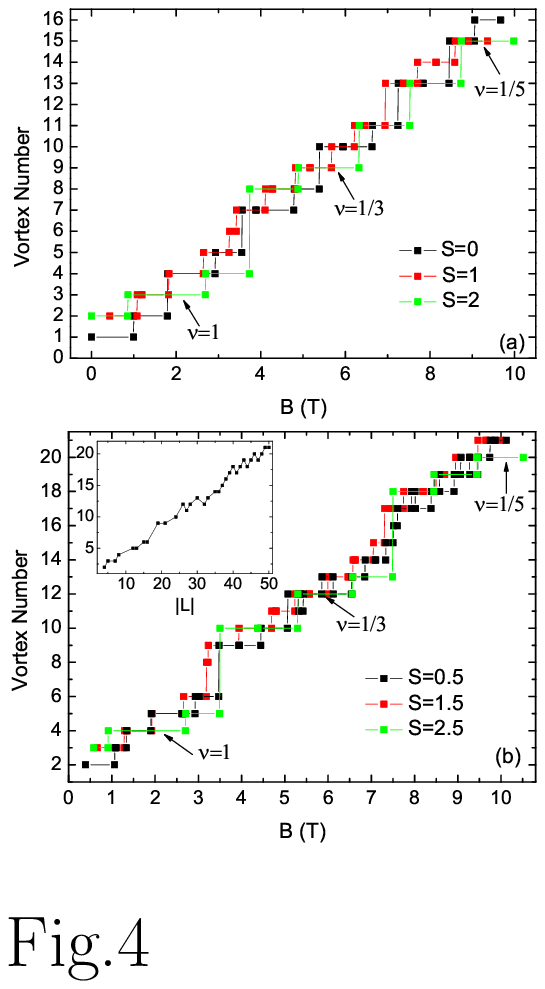}
\caption{\label{FIG:vformation} (Color online) The vortex numbers of
the lowest states with different total spins as functions of the
magnetic fields. For four-electron case (a), $S_z=0$ and the black,
red and green lines correspond to the states with $S=0,1\text{ and
}2$, respectively. For five-electron case (b), $S_z=0.5$ and the
black, red and green lines correspond to the states with
$S=0.5,1.5\text{ and }2.5$, respectively. The vortex numbers of the
states with $S=0.5$ as function of the angular momenta are shown in
the inset for clarity.}
\end{figure}

\begin{figure*}[ht!]
\includegraphics*[angle=0,width=0.72\textwidth]{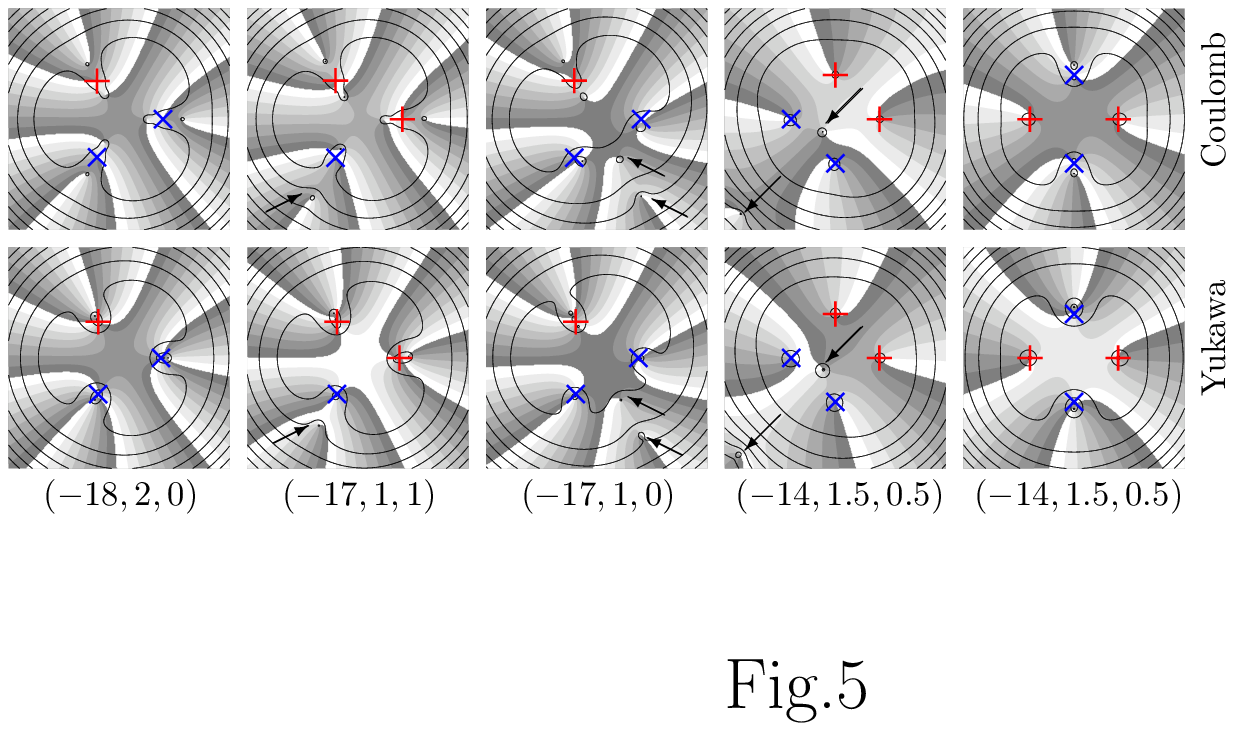}
\caption{\label{FIG:vor4} (Color online) Spin-dependent vortex
distributions of several states with long-range Coulomb (first row)
and Yukawa screened interaction (second row). $+$ and $\times$
indicate the positions of pinned spin-up and -down electrons,
respectively. The arrows indicate the separated vortices.}
\end{figure*}

With the increase of the field, there are also respective angular
momentum transitions for different spin states. The angular momentum
rules for the allowable states which can emerge in the transition
sequences can be obtained from the electron molecules theory in
strong fields\cite{Maksym1995, Maksym2000} or the exact
diagonalization. The exact diagonalization also reveals that some
states whose angular momenta are in accordance with the rules can be
absent from the transition sequence when the electronic states are
still liquidlike. With the transition from the liquid to crystal
states, the absences gradually disappear. For four- and
five-electron states, the transition rules and the absent states are
listed in Appendix. In the four-electron case, only two states are
absent from the transition sequences because the magnetic field can
easily make the electrons form rotating Wigner molecules with such a
small particle number. However, in the five-electron case, the
absences are much more and only disappear in strong fields.

In the transitions, the absolute values of the angular momenta
increase. The vortices also gradually increase. We illustrate the
vortex numbers of the lowest states with different $S$ and the
lowest $S_z$ in Fig.\ref{FIG:vformation}. The complete data are
listed in the Appendix. In the counts, the vortex numbers are those
displayed by the conditional wave functions and the vortices which
move to the infinity with screened interaction are excluded. The
vortex numbers of full-polarized states increase monotonously. And
although there are degenerations due to different $S_z$, the
vortices of full-polarized states with different $S_z$ have no
differences. So in the following discussions we will focus on the
states with lower $S$. It can be seen that the vortex numbers in the
transition sequences of the non-full-polarized states are monotone
non-decreasing when the magnetic field is not very strong, for
five-electron case $\nu\gtrsim 1/3$, i.e. $|L|\lesssim30$. Such
monotonicity is not preserved when the field becomes strong,
especially the five-electron case, see also the inset of
Fig.\ref{FIG:vformation}b where the vortex numbers as the function
of $|L|$ are shown for clarity. By inspecting the vortices of all
the states whose angular momenta are in accordance with the
transition rules, it is found that the vortex numbers of
five-electron states with $L=6,11,14,20,23,31$ and $S=0.5$ exceed
those of the next neighbor ones. And they are absent from the
transition sequence to avoid the breakdown of the monotonicity of
the vortex number with respect to the field or the angular momentum.
There are also other absent states whose total vortex numbers do not
break the monotonicity, in the next subsection, we will analyze the
vortex distributions and discuss the absences more in-depth.

\subsection{Insight of the vortex formation}
In 2DEGs, if the spin degree of freedom is taken into account, when
the magnetic field deviates from the values corresponding to the
fractional filling factors, the total spin of the ground state is no
longer full polarized. Along with the changes of the total spin,
there are also quasi-particle (quasi-electron or -hole) excitations,
namely the reversed-spin quasi-particle excitations\cite{Oaknin1996,
Szlufarska2001} or skyrmions. These excitations will change the
many-body wave functions and be reflected in the formations and
redistributions of vortices. However, due to the long range Coulomb
interaction, the vortices in QDs are dispersed. Especially when the
electrons form the RWMs states, the vortices will spread over the
whole area. So we employ the Yukawa screened Coulomb
interaction\cite{Stopa2006} to clearly show the origins and
behaviors of different vortices in QDs.

In Fig.\ref{FIG:vor4} we show the vortex distributions of some
states as examples. First kind of states is the ones corresponding
to the fractional filling factor $1/(2p+1)$, like the four-electron
state with $L=-18$ and $S=2$ whose filling factor is $1/3$. Such
states with different $S_z$ have same vortex distribution as shown
in Fig.\ref{FIG:vor4} where the state (-18,2,0) is taken as an
example. By the consideration of the screened Coulomb interaction,
the dispersive vortices can approach the positions of electrons.
This is just the scenario of the Laughlin limit in 2DEGs.

When the field deviates from the value corresponding to the
fractional filling factors, the states with lower spins can become
the ground states. Along with such reversion of the total spin of
the ground state, another kind of the vortices can be identified. As
illustrated in Fig.\ref{FIG:vor4}, in the states (-17,1,1) and
(-17,1,0), there are not only the vortices similar to those in
(-18,2,0) but also the separated vortices which do not approach the
positions of the electrons when the interaction is screened, as
indicated by arrows in the plots. Such vortices, which keep
separated from others and do not move to the infinity when the
screened interaction is considered, are analogy of the
quasi-particle in 2DEGs. An interesting feature is that the two
states have different vortex distributions although they are
degenerate in energy and have same total spin. In
Fig.\ref{FIG:vor4}, it can be seen that the vortex numbers of
(-17,1,1) and (-17,1,0) are same as that of (-18,2,0) because they
have same filling factor. For (-17,1,1), there is one separated
vortex. The state (-17,1,0) has one more separated vortex than
(-17,1,1). The reason is that the two identical spin-down electrons
in (-17,1,0) must have same vortex number. So one of the vortices
belonging to a spin-up electron of (-17,1,1) must leave the electron
when it becomes spin-down. Similar differences of vortex
distributions also exist in some other degenerate states with
different $S_z$. In fact, these degenerate states also have
different entanglement entropies, because both the vortex
distribution and the entropy reflect the component differences
between the states.

For five-electron case, there are also the separated vortex
excitations. However, when we recognize the separated vortices by
employing the screened interaction, it must be realized that the
display of the behavior of a vortex may depend on the fixation
manner of the pinned electron. The merging behavior of a vortex to
the position of the electron may be only the result of the
inspection from a particular `viewpoint'. That is to say, only those
merging behaviors irrespective to the fixation manner are true. We
present an example in Fig.\ref{FIG:vor4}. For the five-electron
state (-14,0.5,0.5), the total vortex number in the plots is six. If
the pinned electrons are in alternative mode, it seems that there
are two vortices which can approach the position of the spin-down
electron. But from the viewpoint of the neighbor mode, only one
vortex does so, another one keeps separated from the electron. Then
we can realize that the merging behavior is only a false expression
due to the particular `viewpoint' and there are two separated
vortices in (-14,0.5,0.5).

By making great effort on the classification of the vortices, now we
can return to the analysis of the absent states in the angular
momentum transitions. The behavior of the separated vortices implies
that they have no use in reducing the short-range interaction
between electrons. We know that the concentrated vortices in the
Laughlin wave function are most efficient way to reduce the
short-range interaction. Then we see that when the short-range
interaction becomes more and more important, there are vortices in
the quantum dot which approach the Laughlin limit to reduce the
interaction except the separated ones. From the tables in Appendix,
we can find that most of the absent states have more separated
vortices than neighbor states. For example, the separated vortex
numbers of the four-electron absent state (-6,0,0) and all the
five-electron absent states with $S=0.5$ exceed those of the
neighbor states. Most of these absent states have no less than three
separated vortices. In fact, more separated vortices make the vortex
distribution of those states more dispersed than the neighbor ones
and are unfavorable in energy when the short-range interaction is
still important. Then those states may be absent in the transition
sequences for the lowest states. The states with $S$ unequal to the
lowest value are more complicated. As discussed in
Fig.\ref{FIG:vor4}, the states with same $L$,$S$ but different $S_z$
can have different separated vortices. However, within the states
with same $S_z$ it can still be found that the absent states at
least have more separated vortices than the neighbor states with
smaller $|L|$. The only one special absent state is (-17,1.5,0.5).
It has fewer vortices than the neighbor states and so few vortices
also have no advantage in reducing the energy.

As mentioned previously, with the increase of the magnetic field,
the electrons gradually form the rotating Wigner molecules. The
short-range interaction becomes unimportant and the difference
between separated and other vortices can be ignored. Then even the
states with more separated vortices are no longer unfavorable in
energy and the absences of the states in the transition sequences
gradually disappear. Besides this, the monotonicity of the vortex
number in the angular momentum transition need not to be preserved.
The data in the tables in Appendix also show these conclusions.

\section{Summary}
In summary, we have investigated the vortex structures of the
electronic states in quantum dot without the Zeeman splitting. The
vortex display with spin degree of freedom depends on both the
choice of the probe electron and the fixation manner of the pinned
electrons in the conditional single-particle wave function. By
choosing an appropriate way to fix the pinned electrons in the
functions we explore the transition patterns of the vortex number in
magnetic fields for the lowest states with different spins. It is
found that the vortex number increases monotonously when the field
is not very strong and the states with certain angular momenta are
absent from the transition sequences. When the field becomes strong,
the absences disappear and the monotonicity may not be preserved. By
examining the behaviors of the vortices with different range of the
interaction between electrons, we can identify two kinds of vortices
in the quantum dot which are respectively analogous to the vortices
of the electrons and reversed-spin quasi-particles in the fractional
quantum Hall system. The quasi-particle-like vortices do not
approach the position of the electron when the interaction is
screened and have no use in reducing the short-range interaction.
Then the states with more such separated vortices are unfavorable in
energy when the field is not very strong and may be absent from the
transition sequences. These results imply that the understanding of
the vortex structures and formations with spin degree of freedom are
important for the studies and the control of the spin-relevant
electronic states in nanostructures in magnetic fields.

\begin{acknowledgments}
Financial supports from NSF China (Grant No. 10574077), the``863"
Programme of China (No. 2006AA03Z0404) and the ``973" Programme of
China (No. 2005CB623606) are gratefully acknowledged.
\end{acknowledgments}

\appendix
\section*{Appendix: Vortex for the lowest state with different spins}
The vortex distributions with respect to all allowable angular
momenta for the lowest states with different total spins are listed
in the Tab.\ref{TAB:VorN4} and Tab.\ref{TAB:VorN5}. The columns
'total' are the total vortex numbers displayed by the conditional
single-particle wave functions. The following three columns are
respectively the vortices of per spin-up, per spin-down pinned
electron and the separated vortices identified by the screened
interaction.
\begin{table}[h!]
\caption{Vortex numbers of the four-electron states with $S=0$ and 1
and $S_z=0$. The states marked with $*$ are those absent from the
transition sequences according to the results of the exact
diagonalization.}\label{TAB:VorN4}
\begin{tabular}{|c|c|c|c|c||c|c|c|c|c|}
\hline \multicolumn{5}{|c||}{$S=0$}&\multicolumn{5}{c|}{$S=1$}\\
\hline &\multicolumn{4}{c||}{Vortex Number}&&\multicolumn{4}{c|}{Vortex Number}\\
\cline{2-5}\cline{7-10}\raisebox{1ex}[0pt]{$|L|^\dagger
$}\footnote[0]{$^\dagger$ $L$ can be arbitrary even integer for the
state with $S=0$}&total&per $\uparrow$&per
$\downarrow$&sep.&\raisebox{1ex}[0pt]{$|L|^\ddagger$}\footnote[0]{$^\ddagger$
$L$ can be arbitrary integer but unequal to $4n+2$ for the state
with $S=1$}&total&per $\uparrow$&per $\downarrow$&sep.\\
\hline 2&1&1&0&0&3&2&1&0&1\\
\hline 4&2&1&0&1&4&3&1&0&2\\
\hline 6*&4&1&0&3&5&3&1&0&2\\
\hline &&&&&7&4&1&1&1\\
\hline 8&4&1&1&1&8*&5&1&1&2\\
\hline 10&5&1&2&0&9&5&1&1&2\\
\hline &&&&&11&6&1&1&3\\
\hline 12&7&1&2&2&12&7&1&1&4\\
\hline 14&7&3&2&0&13&8&1&1&5\\
\hline &&&&&15&8&3&2&1\\
\hline 16&8&3&2&1&16&9&3&2&2\\
\hline 18&10&3&3&1&17&9&3&2&2\\
\hline &&&&&19&10&3&3&1\\
\hline 20&10&3&3&1&20&11&3&3&2\\
\hline 22&11&3&4&0&21&11&3&3&2\\
\hline &&&&&23&13&3&3&4\\
\hline 24&13&3&3&4&24&13&3&3&4\\
\hline 26&13&5&4&0&25&14&4&4&2\\
\hline &&&&&27&14&5&4&1\\
\hline 28&15&5&4&2&28&15&5&4&2\\
\hline 30&16&5&5&1&29&15&5&4&2\\
\hline
\end{tabular}
\end{table}

\begin{table}[h!]
\caption{Vortex numbers of the five-electron states with $S=0.5$ and
1.5 and $S_z=0.5$. The states marked with $*$ are those absent from
the transition sequences according to the results of the exact
diagonalization.}\label{TAB:VorN5}
\begin{tabular}{|c|c|c|c|c||c|c|c|c|c|}
\hline \multicolumn{5}{|c||}{$S=0.5$}&\multicolumn{5}{c|}{$S=1.5$}\\
\hline &\multicolumn{4}{c||}{Vortex Number}&&\multicolumn{4}{c|}{Vortex Number}\\
\cline{2-5}\cline{7-10}\raisebox{1ex}[0pt]{$|L|^\dagger
$}\footnote[0]{$^\dagger$ $L$ can be arbitrary integer for the state
with $S=0.5$}&total&per $\uparrow$&per
$\downarrow$&sep.&\raisebox{1ex}[0pt]{$|L|^\ddagger$}\footnote[0]{$^\ddagger$
$L$ can be arbitrary integer but unequal to $5n$ for the state
with $S=1.5$}&total&per $\uparrow$&per $\downarrow$&sep.\\
\hline 4&2&1&0&0&4&2&1&0&0\\
\hline 5&3&1&0&1&&&&&\\
\hline 6*&4&1&0&2&6&3&1&0&1\\
\hline 7&3&1&0&1&7*&4&1&0&2\\
\hline 8&4&1&0&2&8&4&1&0&2\\
\hline 9*&5&1&0&3&9&4&1&0&2\\
\hline 10*&5&1&0&3&&&&&\\
\hline 11*&6&1&0&4&11&5&1&1&1\\
\hline 12&5&1&1&1&12*&6&1&1&2\\
\hline 13&5&1&1&1&13*&6&1&1&2\\
\hline 14*&7&1&1&3&14&6&1&1&2\\
\hline 15&6&1&1&2&&&&&\\
\hline 16&6&1&1&2&16&8&1&1&4\\
\hline 17*&7&1&1&3&17*&7&1&1&3\\
\hline 18*&8&1&1&4&18&9&1&2&3\\
\hline 19&9&1&2&3&19*&10&1&2&4\\
\hline 20*&10&1&2&4&&&&&\\
\hline 21&9&1&2&3&21*&10&1&2&4\\
\hline 22*&10&1&2&4&22&10&1&2&4\\
\hline 23*&11&1&2&5&23&11&3&2&1\\
\hline 24&10&3&2&0&24*&11&1&1&7\\
\hline 25*&11&2&2&3&&&&&\\
\hline 26&12&3&2&2&26&11&3&2&1\\
\hline 27&11&3&2&1&27&12&3&2&2\\
\hline 28&12&3&2&2&28&12&3&2&2\\
\hline 29*&13&3&2&3&29&12&3&2&2\\
\hline 30&13&3&3&1&&&&&\\
\hline 31*&14&3&3&2&31&13&3&3&1\\
\hline 32&12&3&3&0&32&13&3&3&1\\
\hline 33&13&3&3&1&33&14&3&3&2\\
\hline 34*&14&3&3&2&34&14&3&3&2\\
\hline 35&14&3&3&2&&&&&\\
\hline 36&14&3&3&2&36&15&3&3&3\\
\hline 37&15&3&4&1&37&17&4&3&3\\
\hline 38&16&4&3&2&38&17&3&3&5\\
\hline 39&17&3&4&3&39&18&4&4&2\\
\hline 40&18&3&4&4&&&&&\\
\hline 41&17&3&4&3&41&18&4&4&2\\
\hline 42&18&4&4&2&42&18&3&4&4\\
\hline 43&19&3&5&3&43&19&4&4&3\\
\hline 44&18&5&4&0&44&19&4&4&3\\
\hline 45&19&4&4&3&&&&&\\
\hline 46&20&5&4&2&46&20&5&4&2\\
\hline 47&19&4&4&3&47&20&5&4&2\\
\hline 48&20&5&4&2&48&21&5&5&1\\
\hline 49&21&5&4&3&49&21&5&5&1\\
\hline 50&21&5&5&1&&&&&\\
\hline
\end{tabular}
\end{table}


\end{document}